\documentclass[preprints,review,accept,moreauthors,pdftex,10pt,a4paper]{mdpi} 

\newcommand{\vek}{\mathbf}
\usepackage{textcomp}

\firstpage{1} 
\pubvolume{xx}
\issuenum{1}
\articlenumber{5}
\pubyear{2019}
\copyrightyear{2019}
\history{Received: date; Accepted: date; Published: date}
\Title{ The Unique Blazar OJ~287 and its Massive Binary Black Hole Central Engine}

\Author{Lankeswar Dey$^{1}$*\orcidA{}, Achamveedu Gopakumar$^{1}$, Mauri Valtonen$^{2,3}$, Stanislaw Zola$^{4,5}$\orcidB{}, Abhimanyu Susobhanan$^{1}$, Rene Hudec$^{6,7}$, Pauli Pihajoki$^{8}$, Tapio Pursimo$^{9}$, Andrei Berdyugin$^{3}$, Vilppu Piirola$^{2}$, Stefano Ciprini$^{10,11}$, Kari Nilsson$^{2}$, Helen Jermak$^{12}$, Mark Kidger$^{13}$ and Stefanie~Komossa$^{14}$}

\AuthorNames{Lankeswar Dey, A. Gopakumar, M. Valtonen et al.}

\address{%
$^{1}$ \quad Department of Astronomy and Astrophysics, Tata Institute of Fundamental Research, Mumbai 400005, India; gopu@tifr.res.in (A.G.); s.abhimanyu@tifr.res.in (A.S.)\\
$^{2}$ \quad Finnish Centre for Astronomy with ESO, University of Turku, FI-20014, Turku, Finland; mvaltonen2001@yahoo.com (M.V.); piirola@utu.fi (V.P.); kani@utu.fi (K.N.)\\
$^{3}$ \quad Department of Physics and Astronomy, University of Turku, FI-20014, Turku, Finland; andber@utu.fi (A.B.)\\
$^{4}$ \quad Astronomical Observatory, Jagiellonian University, ul. Orla 171, Cracow PL-30-244, Poland; szola@oa.uj.edu.pl (S.Z.)\\
$^{5}$ \quad Mt. Suhora Astronomical Observatory, Pedagogical University, ul. Podchorazych 2, PL30-084 Cracow, Poland\\
$^{6}$ \quad Czech Technical University in Prague, Faculty of Electrical Engineering, Technicka 2, Prague 166 27, Czech Republic; rene.hudec@gmail.com (R.H.)\\
$^{7}$ \quad Engelhardt Astronomical observatory, Kazan Federal University, Kremlyovskaya street 18, 420008 Kazan, Russian Federation\\
$^{8}$ \quad Department of Physics, University of Helsinki, Gustaf H\"allstr\"omin katu 2a, FI-00560, Helsinki, Finland; pauli.pihajoki@helsinki.fi  (P.H.)\\
$^{9}$ \quad Nordic Optical Telescope, Apartado 474, E-38700 Santa Cruz de La Palma, Spain; tpursimo@not.iac.es (T.P.) \\
$^{10}$ \quad Space Science Data Center - Agenzia Spaziale Italiana, via del Politecnico, snc, I-00133, Roma, Italy; stefano.ciprini.asdc@gmail.com (S.C.) \\
$^{11}$ \quad Instituto Nazionale di Fisica Nucleare, Sezione di Roma Tor Vergata, Roma, I-00133, Italy \\
$^{12}$ \quad Astrophysics Research Institute, Liverpool John Moores University, IC2, Liverpool Science Park, Brownlow Hill, L3 5RF, UK;  h.e.jermak@ljmu.ac.uk (H.J) \\
$^{13}$ \quad Herschel Science Centre, ESAC, European Space Agency, 28691 Villanueva de la Ca{\~n}ada, Madrid, Spain; mkidger@sciops.esa.int (M.K) \\
$^{14}$ \quad Max Planck Institut fuer Radioastronomie, Auf dem Huegel 69,53121 Bonn, Germany; astrokomossa@gmx.de (S.K.) \\

}

\corres{Correspondence: lankeswar.dey@tifr.res.in; Tel.: +91-22-2278-2428}

\abstract{
The bright blazar OJ~287 is the best-known candidate for hosting a nanohertz gravitational wave (GW) emitting supermassive binary black hole (SMBBH) in the present observable universe.
The binary black hole (BBH) central engine model, proposed by Lehto and Valtonen in 1996, was influenced by the two distinct periodicities inferred from the optical light curve of OJ~287.
The current improved model employs an accurate general relativistic description to track the trajectory of the secondary black hole (BH) which is crucial to predict the inherent impact flares of OJ~287.
The successful observations of three predicted impact flares open up the possibility of using this BBH system to test general relativity in a hitherto unexplored strong field regime. 
Additionally, we briefly describe an on-going effort to interpret 
observations of OJ~287 in a Bayesian framework.
}

\begin{document}

\section{Introduction}
\label{sec:intro}
It is now well established that nearly all massive active and normal galaxies contain supermassive black holes (SMBHs) at their centers \cite{kormendy95, ff05, graham16}. 
The SMBH of a typical active galaxy is surrounded by an accretion disk and accretion on to the massive BH fuels the system.
The accretion-induced luminosity, arising from its small central region, can be comparable to, often exceeds, the luminosity of the rest of the galaxy.
These central regions are usually referred to as active galactic nuclei (AGNs) \cite{kembhavi99}. 
Some AGNs also launch relativistic jets, beams of relativistic ionized matter, along the axis of rotation. If the direction of the jet of an AGN is along our line of sight, we call the AGN a blazar. The radiation we see from a blazar is dominated by the emission from the jet.

As the mergers of galaxies appear to be frequent, it is expected that at the center of some of the galaxies, instead of one SMBH there can be two of them. 
A number of astrophysical considerations point to the possibility that mergers of two galaxies can lead to the formation of gravitationally bound SMBH binaries \cite{begelman80,MM05}.
While a number of observations have been considered as signatures of SMBH pairs at wide and small separation \cite{komossa16}, the blazar OJ~287 (at redshift $z$=0.304) is the best candidate for hosting a binary SMBH at its center \cite{lv96, dey18}.
We list below various observational and theoretical pieces of evidence 
that strongly point to the presence of an SMBH binary in the central engine of OJ~287.

\begin{figure}[t]
\centering
\includegraphics[width=0.9\textwidth]{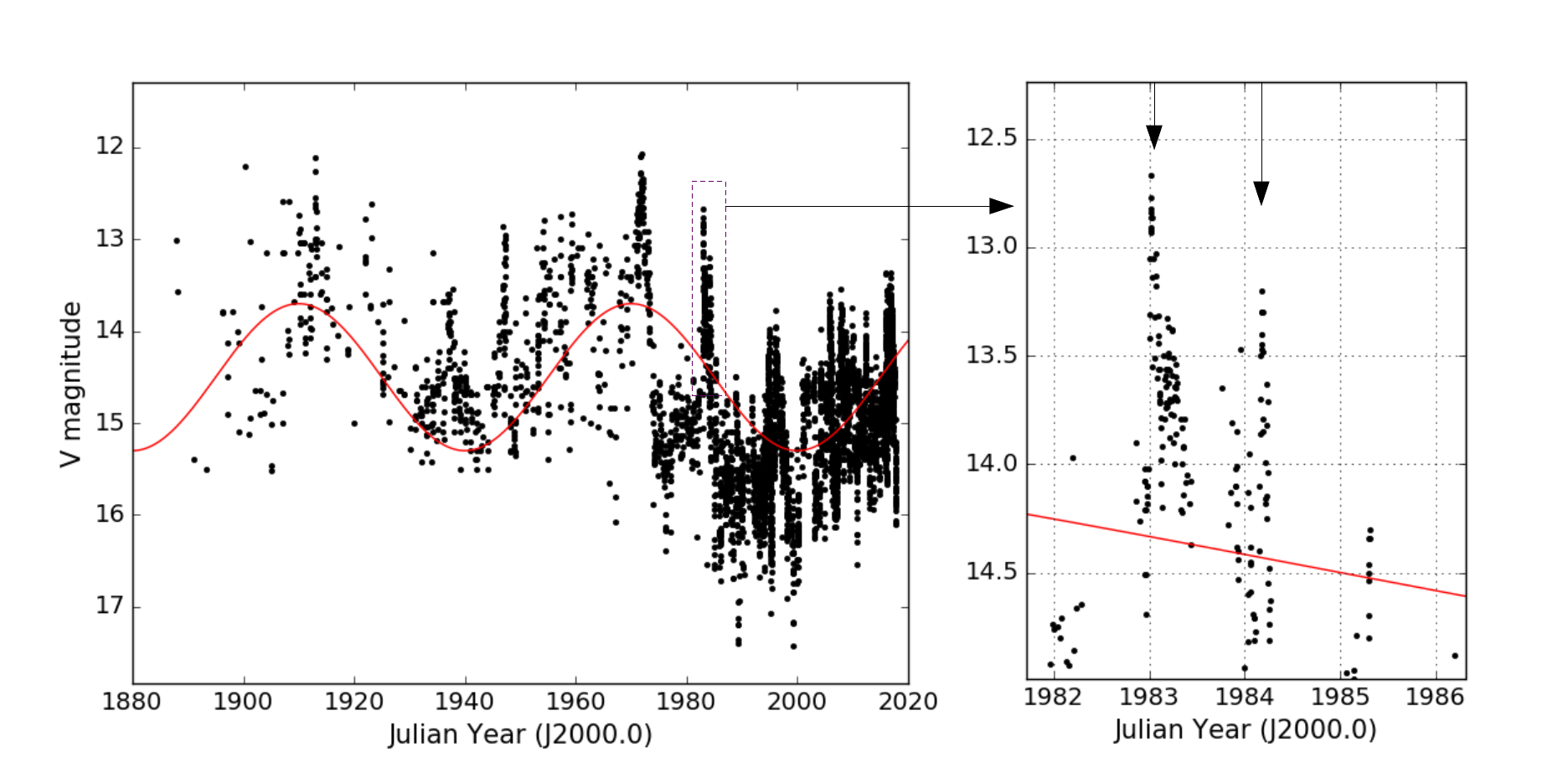}
\caption{ The left panel shows the optical LC of OJ~287 from 1888 to 2018. The red sinusoidal curve helps to visualize the longer 60 year periodicity in the LC. 
The right panel focuses on the LC from 1982 to 1986 for demonstrating the doubly peaked nature of these outbursts \cite{dey18}.}
\label{fig:light_curve}
\end{figure}

During the 1980s,  the Tuorla Observatory started a quasar monitoring program and OJ~287 grabbed attention due to its quasi-periodic doubly-peaked outburst pattern in its optical light curve (LC) \cite{val88,sillanpaa88}. In Figure~\ref{fig:light_curve} \cite{dey18}, we show the optical LC of OJ~287 which goes all the way back to the year 1888!
This data set exists due to the proximity of OJ~287 to the ecliptic and therefore was unintentionally photographed often in the past, providing us with the LC of OJ~287 extending back to $\sim 130$ years. 
A visual inspection of the LC reveals the presence of two periodic variations with approximate timescales of 60 and 12 years, which have been confirmed through a detailed quantitative analysis \cite{val06a}. 
In the left panel of Figure~\ref{fig:light_curve}, we mark the 60 year period by a red sinusoidal curve. Additionally, it is possible to infer the presence of regular pairs of outbursts at $\sim 12$ years interval, where the two peaks are separated by a few years in the LC (see the right panel of Figure~\ref{fig:light_curve}).
The presence of such a double periodicity in the optical LC provided possible evidence for the occurrence of a quasi-Keplerian orbital motion in the blazar.
In this description, the 12-year periodicity corresponds to the orbital period timescale and the longer 60-year timescale is associated with the advance of periastron.

\begin{figure}[H]
\centering
\includegraphics[width=0.9\textwidth]{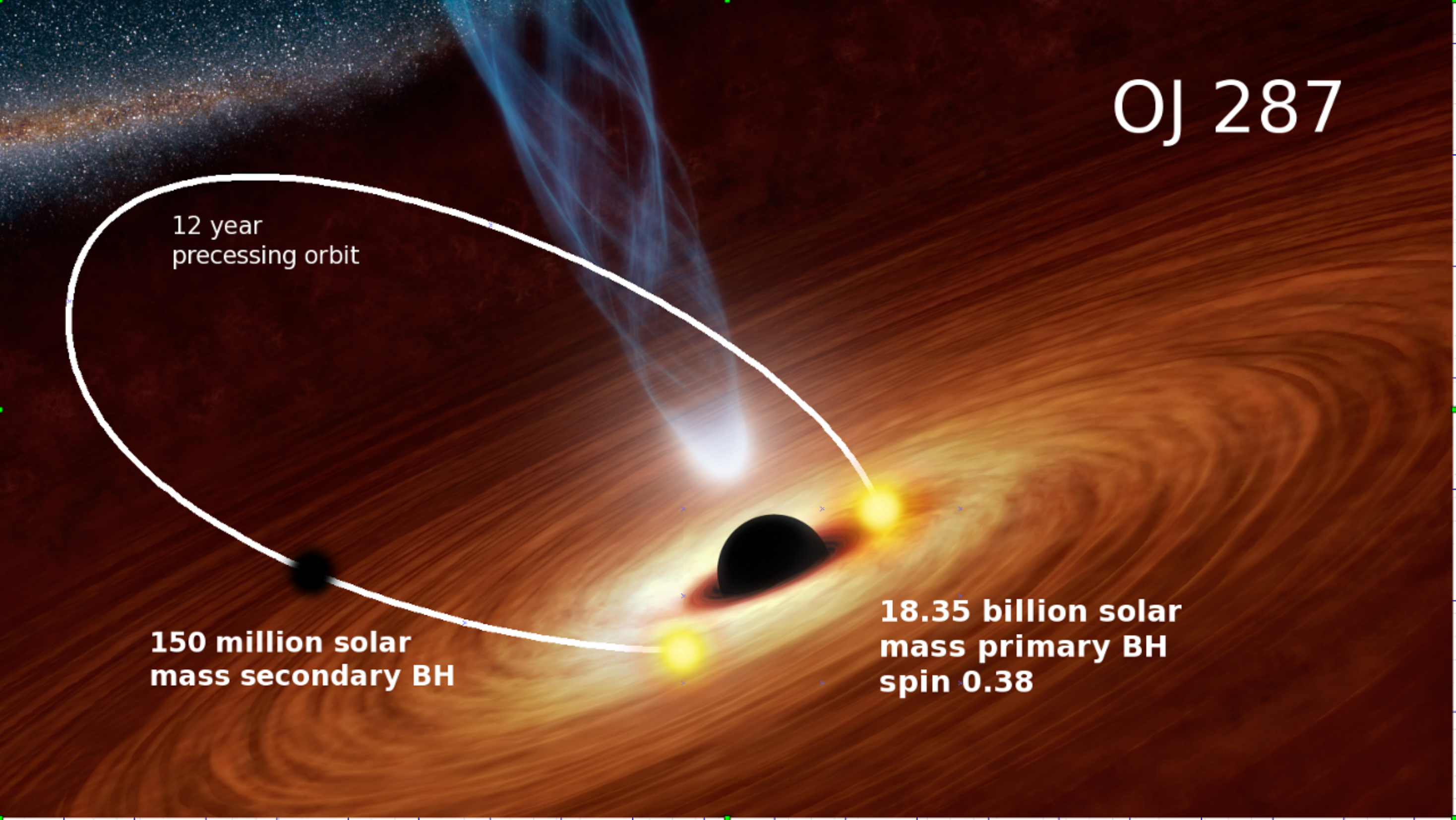}
\caption{Artistic illustration of the binary black hole system in OJ~287 \cite{dey18}.\textcopyright AAS 2018}
\label{fig:oj_model}
\end{figure}

Therefore, a possible model to explain the observed periodicities in the LC naturally involves a secondary black hole that orbits a more massive primary BH in an eccentric orbit \cite{lv96,sun97}.
Additionally,  the secondary BH impacts the accretion disk of the primary BH twice during one orbit having a (redshifted) period of $ \sim 12$ years (Figure~\ref{fig:oj_model}). 
These two impacts produce the two flares that are observed more than a year apart, and such impact flares are repeated during every orbital cycle. 
We invoke a simple prescription to model 
the astrophysical process that generates these impact flares. The impact of the secondary BH with the accretion disk releases hot bubbles of gas from the disk on both sides \cite{ivanov98, pihajoki16}. These hot bubbles then expand and cool down until they become optically thin, and the radiation from the entire volume is seen. The radiation is thermal bremsstrahlung at a temperature of $ \sim 10^5$ K \cite{val12}.

A number of alternative models, available in Refs.~ \citet{katz97,tanaka13, britzen18}, did try to explain the observed variations in the LC of OJ~287 that include e.g. Doppler boosting variation from a turning jet, varying accretion rate, etc. 
However, the following additional observed features are not explained by other models:
\begin{itemize}[leftmargin=*,labelsep=5.8mm]
\item    The flares rise rapidly with the rise time of only about a few days. In contrast, the timescales associated with processes like jet turning are several months to years.
\item    At the times of impact flares, the degree of polarization always goes down \cite{val08b, val17}. 
In the BBH model, this is due to an additional unpolarized component associated with the bremsstrahlung radiation at the time of outburst, which is in addition to the long-lived jet emission.
\item     A natural (and a powerful) feature of the BBH impact model is its accurate predictive power.
Indeed, the model accurately predicted the starting times for the widely observed 1995, 2005, 2007 and 2015 outbursts \cite{lv96, val08b, val16, val17}.
It is rather difficult to make testable predictions using the alternative models for OJ~287
as they assume that the flares are strictly periodic which puts the prediction off by up to several years. Random deviations around the strictly periodic times do not suit either, as the deviations are predictable.
\end{itemize}
More details about the alternative models can be found in the Appendix.

Many aspects of the BBH impact model, detailed in  \citet{lv96, sun97}, were improved in the subsequent years. 
These developments include better descriptions for the astrophysical processes causing impact outbursts \cite{val07,pihajoki13a}, 
accurate general relativistic orbital description \cite{dey18} as well as the addition of a large number of archived impact outburst data sets \cite{hud13}.
At present, the BBH dynamics is described using the post-Newtonian approximation to general relativity that incorporates the effects
of periastron advance, black hole spin, and GW emission. 
This detailed prescription makes it possible to test general relativity in strong field regimes not tackled by  
relativistic binaries present in   PSR J0737-3039, LIGO/VIRGO GW events, and stars orbiting the central massive BH of our galaxy \cite{wex14,abbott16,lvc18, hees17}.
Additional investigations are also being pursued to substantiate the presence of a $1.8$ billion solar mass BH in OJ~287.
This includes an effort to verify from observations whether the host galaxy of OJ~287 follows the established BH mass-bulge luminosity correlation \cite{kb11}
and employing BH binary scenario to model temporal variations of the jet position angle using high-frequency radio data sets \cite{val12b,val13,dey19}.

In Section~\ref{sec:develop_BBH}, we summarize various developments implemented into the original BBH model and the current state of the model. Section~\ref{sec:testGR} describes various tests of general relativity being pursued or possible with the BBH central engine model for OJ~287.
A new accurate and efficient approach to track the orbit of the secondary BH and to extract astrophysical information from observations are discussed in Section~\ref{sec:phasing}. 
A brief summary and possible future directions are listed in Section~\ref{sec:discussion}.

%%%%%%%%%%%%%%%%%%%%%%%%%%%%%%%%%%%%%%%%%%
\section{ Construction of the BBH central engine model for OJ~287}
\label{sec:develop_BBH}

This section provides a brief step-by-step description of the BBH central engine model. 
We begin by showing how fairly basic astrophysical considerations can be invoked to constrain the important parameters of our model. However, we need to invoke the observed outburst timings for accurately estimating various parameters of the BBH impact scenario for OJ~287.
How to employ outburst timings and a general relativistic description of the BBH for the above purpose is listed in subsections~\ref{subsec:impact_details} and \ref{subsec:pn_model}.
This is followed by the current description of our impact flare model for OJ~287.

\subsection{ Elements of the BBH model from astrophysical considerations}
\label{subsec:estimate_params}
The basic scenario involves a secondary BH that orbits a more massive primary BH and the outbursts occur when the secondary BH impacts the accretion disk of the primary. 
Interestingly, this scenario allows us to estimate the values of many important model parameters by invoking a few astrophysical considerations. 
The major flare epochs provide a time sequence of events. It was realized early on that this sequence can be generated by a simple, purely mathematical rule (\citet{lv96} or  model 1). 
This rule requires us to consider a particle moving in a Keplerian elliptical orbit, and let the ellipse rotate forward by a constant rate. This leads to a sequence of times when the particle crosses a fixed line in the orbital plane, drawn through the Keplerian focal point of the ellipse. For every value of eccentricity and rotation rate of the ellipse, a time sequence is created and in general, such time sequences need not have anything to do with OJ~287 observations.
However, if we choose  eccentricity and precession rate to be $\sim 0.7$ and $\sim 39$ degrees per orbit, respectively, we get a sequence of epochs that matches fairly well with the OJ~287 flare timings.
The physical and astrophysical implications of the above conclusion was explored in \citet{lv96} ( model 2).
Clearly, the crossing of the line can be naturally associated with the crossing of an inclined plane in a binary system. 
Further, what actually generates the signal is not of great importance as far as the determination of the orbit is concerned \citep{pietila98}. 
Importantly, if we invoke General Relativity (GR) to explain the forward precession rate, we get a very robust value for the total mass of the underlying system in OJ~287.

Let us now add one more assumption. Assume that the two periodicities (12 and 60 yr) in the light curve are related to the two periodicities in the system, the orbital period and the precession period. It does not matter what the mechanism is.
Naturally, the inferred 60-year periodicity of the LC may be associated with half of the orbital precession period as the precession effects are expected to be symmetric about the accretion disk plane (the orbital precession arises from the post-Keplerian orbital trajectory for the secondary BH).
It is straightforward to estimate the orbital precession rate per orbit in degrees to be $\Delta \phi \sim 12 \times 180/60 \ \ deg = 36 \ \ deg$ where we let the orbital period of the binary to be $\sim 12$ years.
Invoking GR to explain the above $\Delta \phi$ value can lead to an estimate for the total mass of the BBH, provided we have an estimate for the orbital eccentricity.
It is reasonable to use the minimum temporal separation between the two peaks of the doubly-peaked outbursts to constrain the eccentricity. This is the value we get from the purely mathematical model. The ratio of the minimum time interval between the two nearby impact flare peaks and the orbital period should be larger for orbits with small eccentricities. In contrast, this ratio is expected to smaller for eccentricity values close to unity.
It turns out that the optimal value for orbital eccentricity is $\sim 0.7$ from 
observations.
We may now provide an estimate for the total mass ($M$) of the BBH system in OJ~287 with the help of 
the following expressions for the orbital period ($P$) and the precession rate ($\Delta \phi$), namely

\begin{equation}
    P^2 = \frac{4 \pi^2 \, a^3}{G\,M} \,,
\end{equation}
\begin{equation}
    \Delta \phi = \frac{6\pi \, G \,M}{c^2 \, a \, (1 - e^2)} \,,
    \label{eqn:delta_phi}
\end{equation}
where $a$ is the semi-major axis of the orbit. 
With the help of the inferred $P$, $\Delta \phi$ and $e$ estimates, we get the approximate total mass of the system $M \sim 2\times10^{10} M_{\odot}$ after eliminating the parameter $a$ from the above two equations. 
Recall that Equation~\ref{eqn:delta_phi} provides only the leading order GR contributions 
to periastron advance, characterized by $\Delta \phi$.
The higher order corrections provide additional positive contributions to $\Delta \phi$ and therefore the above $M$ estimate is a slight overestimation.
Let us emphasize that no additional astrophysical considerations are invoked to 
obtain the above $M$ estimate.
Additionally, it is possible to provide an estimate for the mass of the secondary BH by employing the astrophysics of the impact and the strength of the outburst signal \cite{ivanov98, val12}. The use of the observed outburst luminosity 
of 5.6 mJy 
for the 2007 outburst and the \citet{lv96}  model lead to an estimate for the secondary BH mass to be $m_{sec} \sim 1.4 \times 10^8 M_{\odot}$ \citep{val12}.
Let us emphasize that these estimates were extracted without really employing the accurate timing data from the observed outbursts from OJ~287, only the first order fit to the Keplerian orbital dynamics with the GR forward precession. 
A more precise estimates of the BBH parameters by explicitly using outburst timings are given in subsection~\ref{subsec:current_status}.

\subsection{ Details of BBH central engine from the timing of impact outbursts }
\label{subsec:impact_details}
We now move on to describe how we estimate various parameters of the BBH central engine model using the outburst timings.
This requires us to bring in additional astrophysical considerations as the outbursts do not take place just at the epochs of secondary BH impacts. 
Such impacts are expected to result in the creation of two hot plasma bubbles which come out from both sides of the accretion disk.
These plasma bubbles expand, cool, and after a time delay, become optically thin. This is when the radiation can escape the gas plasma bubble and manifest as the impact flare.
The time interval taken by these bubbles to become optically thin is termed as the `time delay' and it clearly affects the observed starting epoch of an outburst.
Therefore, accurate modeling of the time delay is crucial for us.

A calculation to compute the above time delay was first provided in \citet{lv96}. 
Subsequently \citet{ivanov98} carried out a hydrodynamic simulation of the impact and the release of a hot cloud of plasma from the disk. Their simulation showed that plasma cloud is released from both sides of the disk, one bubble bursting out to the backward direction relative to the motion of the secondary, the other following the orbit of the secondary. The physical properties of the bubbles were calculated and they matched well with what is expected from simple analytic arguments. 
Unfortunately, \citet{ivanov98} did not carry out their calculation far enough to reach the radiation outburst stage.
However, they estimated that the resulting flare should correspond to the Eddington luminosity of the secondary BH at the maximum brightness as argued in \citet{lv96}.
This calculation incorporates the idea that the secondary BH impact creates a compressed spherical blob of plasma.
The optical outburst occurs when such a blob expands by a factor of $\tau^{4/7}$ where $\tau$ is the initial optical depth and this expansion essentially provides the time delay in terms of the accretion disk and BH parameters \cite{lv96,val19}.
Specifically, the time delay depends on the mass of the secondary BH ($m_{\rm sec}$), the relative velocity of the secondary BH with respect to the accretion disk ($v_{\rm rel}$) at the time of impact and density ($n$) and thickness ($h$) of the disk. 
The expression for the time delay given in \citet{lv96} reads

\begin{equation}
    t_{\rm del} = d \, {m_{\rm sec}}^{26/21} \, {v_{\rm rel}}^{-355/84} \, h^{13/21} \, n^{102/112} \,, 
    \label{eqn:t_del}
\end{equation}
where $d$ is a parameter called `delay parameter' which depends on the accretion disk properties. 
Note that the time delay is a function of the secondary mass. The time delay and therefore the secondary mass is also determined as a part of the orbit solution. This method is in no way related to the secondary mass determination using the brightness of the flare. It is remarkable that the two methods give 
essentially the 
same value within the 10 percent uncertainty. The secondary mass value also agrees with the limit provided by the long term stability requirement. 
It was argued that the disk becomes unstable if the primary versus secondary mass ratio is less than 100 \citep{val12}.
It is customary to adapt the $\alpha_g$ disk, in which the viscosity is proportional to the gas pressure \cite{sakimoto81}. In this model, the properties of the accretion disk are uniquely defined by the central mass, the accretion rate of the primary ($\dot{m}$) and the viscosity parameter ($\alpha$).
Using the disk model, \citet{lv96} calculated the variations in the disk properties like $h$ and $n$ with the distance from the central BH. 
This allows one to obtain reasonable estimates for  $h$ and $n$ of the accretion disk at the impact site, provided the impact site distance  ($R_{\rm imp}$) from the central BH is available.
A general relativistic description for the BBH orbit should be able to provide $R_{\rm imp}$ and $v_{\rm rel}$ at the time of impact and therefore, BBH dynamics essentially provides an estimate for the time delay. 
We need to invoke GR as the BBH of the model experiences a periastron advance of $\sim 39^{\circ}$ per orbit. In comparison, the measured advance of periastron for the double pulsar is $\sim 0.005^{\circ}$ per orbit \cite{kramer06}.
In summary, our Equation~\ref{eqn:t_del} provides an astrophysically relevant estimate for the time delays for different impact epochs, essentially from a general relativistic BH binary description.

\begin{figure}[t]
\centering
\includegraphics[width=0.7\textwidth]{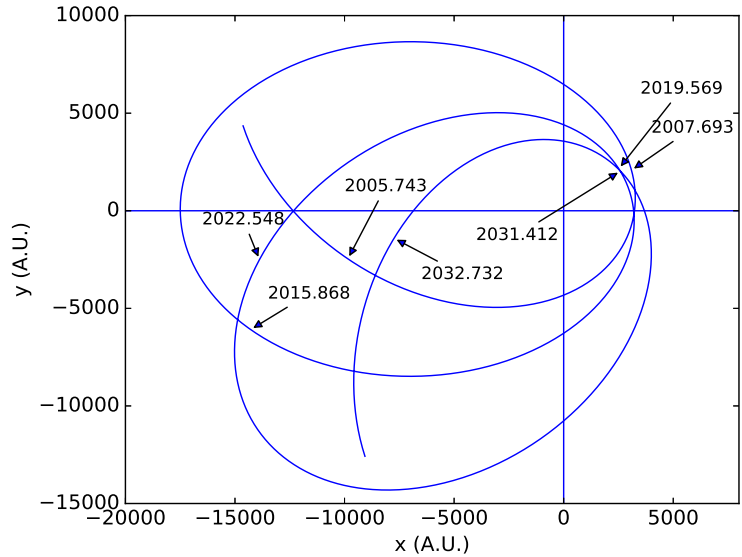}
\caption{The typical orbit of the secondary BH in OJ~287 in 2005--2033 window. 
The primary BH (Schwarzschild radius $R_S \sim 362$ AU) is situated at the origin with its accretion disk in the $y=0$ plane. The average distance of periastron is $\sim 9\, R_S$ and the binary is going to merge within ten thousand years.
 The locations of the secondary BH at the time of different outburst epochs are marked by arrow symbols. The time-delay effect is clearly visible, while close inspection reveals that these delays for different impacts are different  \cite{dey18}. \textcopyright AAS 2018}
\label{fig:orbit}
\end{figure}

An additional astrophysical aspect needs to be incorporated while dealing with the `time delays'. 
The tidal force of the approaching secondary BH causes the accretion disk to be pulled up towards the secondary. 
This forces the impacts to occur at earlier epochs compared to a scenario where we neglect the tidal deformation of the disk and this time difference is termed as `time advance' ($t_{\rm adv}$) (see \citet{val07} for details). 
In practice, we let the accretion disk to lie in a fixed plane as the $y = 0$ plane in Figure~\ref{fig:orbit} \cite{dey18}.
In our approach, the observed starting time of the impact flare is given by 
\begin{equation}
    t_{\rm out} = t_{\rm imp} - t_{\rm adv} + t_{\rm del} \,.
    \label{eqn:t_outburst}
\end{equation}
where $t_{\rm imp}$ stands for the time of the disk impact, obtainable from BBH orbital dynamics.
How we describe the BBH dynamics is explained in the next subsection.

\subsection{  General relativistic orbital trajectory for the secondary BH }
\label{subsec:pn_model}
We employ the post-Newtonian (PN) approximation to GR to describe the evolution of secondary BH trajectory in the strong gravitational field of the primary BH.
The PN approximation provides dynamics of BH binaries as corrections to the Newtonian dynamics in powers of $(v/c)^2 \sim G\,M/(c^2 \, r)$, where $v$, $M$, and $r$ are the characteristic orbital velocity, the total mass, and the typical orbital separation of the binary, respectively. 
In a convenient center of mass frame, the post-Newtonian equations can be written in the form \cite{blanchet14,will17}
\begin{eqnarray}
    \ddot{\vek x} \equiv \frac{d^2{\vek x}} {dt^2} &=& \ddot{{\vek x}}_{\rm 0} + \ddot{{\vek x}}_{\rm 1PN} + \ddot{{\vek x}}_{\rm 2PN} + \ddot{{\vek x}}_{\rm 3PN} \nonumber\\ 
    && + \ddot{{\vek x}}_{\rm 2.5PN} + \ddot{{\vek x}}_{\rm 3.5PN} + \ddot{{\vek x}}_{\rm 4PN(tail)} + \ddot{{\vek x}}_{\rm 4.5PN} \nonumber \\
    && + \ddot{{\vek x}}_{\rm SO} + \ddot{{\vek x}}_{\rm Q} + \ddot{{\vek x}}_{\rm 4PN(SO-RR)} \,, 
\label{eqn:eom}
\end{eqnarray}
where ${\vek x} = {\vek x}_1 - {\vek x}_2$ gives the relative separation vector between the two BHs with masses $m_1$ and $m_2$ in the center-of-mass frame. The familiar Newtonian contribution, denoted by $\ddot{{\vek x}}_{0}$, is given by $\ddot{{\vek x}}_{0} = -\frac{ G\, m}{r^3} \, {\vek x}$, where $m= m_1 + m_2$, $r = |{\vek x}|$. 
Note that nPN contributions provide $\mathcal{O}((v/c)^{2n}) $ correction terms.

The 1PN, 2PN and 3PN terms that appear in the first line of Equation~\ref{eqn:eom} are conservative in nature and essentially force the periastron of the orbit
to precess.
The contributions appearing in the second line of Equation~\ref{eqn:eom},
namely, 2.5PN, 3.5PN, 4PN(tail), and 4.5PN terms,  are due to the emission of GWs from the binary system.
These contributions force the orbital period and eccentricity to decay with time.
\citet{dey18} showed that the contributions of PN terms higher than 4.5 are negligible for the binary dynamics in OJ 287.
The effects of BH spin enter the above equation in the third line and the $\ddot{{\vek x}}_{\rm SO}$ term provides the general relativistic spin-orbit interactions at 1.5PN and 2.5PN orders. 
These terms make the orbital plane of the secondary and spin of the primary to precess while the $\ddot{{\vek x}}_{\rm 4PN(SO-RR)}$ term stands for the spin-orbit radiation reaction terms.
The classical spin-orbit term or the quadrupolar term is denoted by $\ddot{{\vek x}}_{\rm Q}$ and how we plan to test the celebrated {\it Black Hole no-hair theorem} with its help will be discussed in subsecion~\ref{subsec:no-hair}. 
We note that the $\ddot{{\vek x}}_{\rm 4PN(tail)}$ contributions are somewhat different from the other reactive terms.
This term arises due to the scattering of quadrupolar GWs from the space-time curvature created by total mass (monopole) of the system and hence depends, in principle, on the entire history of the system, and are called hereditary contributions \cite{blanchet93}. 
A closed form expression for these contributions do not exist and an effective way of incorporating such contributions is discussed in \citet{dey18}.

We move onto describe how we estimate various parameters of the BBH central engine model for OJ~287 \cite{val07}.
The orbital trajectory of the secondary BH is determined by numerically integrating Equation~(\ref{eqn:eom}).
Initially, we determine the orbit using certain `guesstimates' for various parameters as described in subsection~\ref{subsec:estimate_params}.
From such an orbital description, the times of impacts of the secondary BH with the accretion disk are calculated (these are the times when the orbit crosses the accretion disk plane $y=0$ as in Figure~\ref{fig:orbit}). 
This allows us to obtain the starting times of various outbursts, calculated using Equation~\ref{eqn:t_outburst}. 
Thereafter, we carefully check if these calculated starting times agree with the well-observed outburst epochs within uncertainties.
If not, the parameters are changed and the whole procedure is repeated.
When all the outburst timings, determined by tracking the orbital trajectory of the secondary BH, match with the observed ones, the parameters are taken to be a solution. In Figure~\ref{fig:orbit} we display a typical orbit of the binary system in OJ~287.

\subsection{The most up to date description of the BBH central engine }
\label{subsec:current_status}
Currently, we have accurate measurements for the starting times of ten well-observed outbursts at our disposal to constrain the binary orbit. These ten outburst starting times are given in Table~\ref{tab:outburst_times} with the observational uncertainties \cite{dey18}. 

\begin{table}[H]
\caption{ Extracted starting times of the well observed optical outbursts of OJ 287 \cite{dey18}. \textcopyright AAS 2018}
\centering
%% \tablesize{} %% You can specify the fontsize here, e.g.,  \tablesize{\footnotesize}. If commented out \small will be used.
\begin{tabular}{c}
\toprule
\textbf{Outburst times with estimated uncertainty (Julian year)}\\
\midrule
1912.980    $\pm$  0.020\\ 
1947.283    $\pm$  0.002\\ 
1957.095    $\pm$  0.025\\ 
1972.935    $\pm$  0.012\\
1982.964    $\pm$  0.0005\\  
1984.125    $\pm$  0.01\\ 
1995.841    $\pm$  0.002\\ 
2005.745    $\pm$  0.015\\ 
2007.6915   $\pm$  0.0015\\ 
2015.875    $\pm$  0.025\\ 
\bottomrule
\end{tabular}
\label{tab:outburst_times}
\end{table}

Using these ten outburst times, we determine the binary system parameters using the above-listed procedure. 
These parameters include the mass of the primary BH $m_1$, the mass of the secondary BH $m_2$, the Kerr parameter of the primary BH $\chi_1$, present orbital eccentricity $e_0$, the rate of advance of pericenter per orbit $\Delta \phi$ and the orientation of the semi-major axis $\Theta_0$ in 1856 (starting time of orbit integration). 
Additionally, we can estimate two more parameters related to the effects of astrophysical processes that are associated with the accretion disk impact of the secondary BH. These two parameters are $d$ and $h$ where $d$ is the delay parameter present in Equation~\ref{eqn:t_del}, while the disk thickness parameter $h$ is a scale factor with respect to the ``standard” model of Lehto \& Valtonen (1996). The solutions for these parameters are listed below in Table~\ref{tab:parameters} \cite{dey18}.

\begin{table}[h]
\caption{Parameters of the orbit solution \cite{dey18}. \textcopyright AAS 2018}
\label{tab:parameters}
\centering
%% \tablesize{} %% You can specify the fontsize here, e.g.,  \tablesize{\footnotesize}. If commented out \small will be used.
\begin{tabular}{ccc}
\toprule
\textbf{Parameter}	& \textbf{Value}	& \textbf{Unit}\\
\midrule
 $m_1$ & 18348 & $10^6\, M_{\odot}$ \\
 $m_2$ & 150.13 & $10^6\, M_{\odot}$ \\
 $\chi_1$ & 0.381 & \\
 $e_0$ & 0.657 & \\
 $\Delta\Phi$ & 38.62 & $deg$  \\
 $\Theta_0$ & 55.42  & $deg$   \\
 $h$ & 0.900 & \\
 $d$ & 0.776 & \\
\bottomrule
\end{tabular}
\end{table}

It is possible to derive two additional quantities using the extracted parameters listed in Table~\ref{tab:parameters}.
They are the present orbital period of the binary orbit $P_{\rm orb} = 12.06 \ year$ and the rate of orbital period 
decay $\dot{P}_{\rm orb} = 0.00099$. 
These estimates suggest that the binary BHs will merge within ten thousand years.
Additionally, these accurate estimates allow us to predict the start of the next impact flare outburst to be on July 29, 2019.

%%%%%%%%%%%%%%%%%%%%%%%%%%%%%%%%%%%%%%%%%%
\section{Tests of GR Using OJ~287}
\label{sec:testGR}
Testing Einstein's general theory of relativity in the vicinity of strong gravitational fields is an important endeavor \cite{yunes13}.
% https://link.springer.com/article/10.12942/lrr-2013-9
This is mainly to explore any possible deviations from Einstein's predictions for the behavior of gravitational fields in dynamical and strong gravity scenarios. 
Compact binaries like PSR~B1913+16 pioneered the efforts to test GR in strong field regimes \cite{wex14}.
%https://arxiv.org/abs/1402.5594
Recent observations of GWs from merging BH-BH and Neutron Star- Neutron Star binaries are allowing the test of GR in ultra-strong and dynamical gravitational fields \cite{berti15}.
%https://arxiv.org/abs/1501.07274
It turns out that OJ~287 observations can be used to test GR in the yet-to-be-explored strong filed regime as displayed in the left panel of Figure~\ref{fig:no-hair_correlation}.

\subsection{ First indirect evidence for GW emission from BH-BH binaries}
In Einstein's GR, orbiting compact objects should emit GWs. 
This causes the binary to lose its orbital energy and angular momentum 
which forces the decay of its orbital period and eccentricity.
The existence of GWs from inspiralling NS-NS binaries was demonstrated by the long-term radio timing of PSR~B1913+16 \cite{damour15}.
% https://arxiv.org/pdf/1411.3930
The first indirect evidence for the emission of GWs from orbiting BH-BH binary was provided by the 2007 monitoring of the predicted OJ~287 impact flare \cite{val08b}.
In the BBH central engine model, the secular changes in the secondary BH trajectory can change the computed times of its impact with the accretion disk of the primary BH. 
This can naturally shift the expected starting epochs of the impact flare outbursts.

The indirect evidence for the presence of GW emission in OJ~287 was demonstrated with the help of PN-accurate orbital dynamics, given symbolically by Equation~\ref{eqn:eom}.
We note that the starting time of this outburst was predicted to be September 13, 2007, with an uncertainty of two days, while 
employing the fully 2.5PN accurate equations of motion.
However, the BBH model predicted the start of the 2007 outburst some 20 days later (in early October 2007) when the orbit of the secondary BH was computed without incorporating the 2.5PN order radiation reaction terms in the orbital dynamics.
The 2007 observational campaign for OJ~287, started from September 4, did witness the beginning of the expected outburst on September 12, 2007. 
This crucial observation provided the indirect evidence for the presence of GW emission from the BH-BH system in OJ~287 \citep{val08b}.

\subsection{Testing the BH `no-hair' Theorem during the present decade}
\label{subsec:no-hair}
The celebrated Black Hole `no-hair' theorem states that a rotating BH can be completely described by its mass and angular momentum \cite{NG15}.
%https://journals.aps.org/prl/abstract/10.1103/PhysRevLett.114.151102
This implies that we do not require any equation of state to describe the properties of BHs as we need to do for neutron stars.
In other words, the multipole moments of a rotating BH can be determined in terms of its mass and angular momentum alone (BHs have `no hair') and prompted the formulation of the following test of the theorem \citep{thorne80}.
The test involves estimating the mass, the spin and the quadrupole moment of a rotating astrophysical BH from observations. 
It turns out that the dimensionless quadrupole moment ($q_2$) of a BH in GR is related to its Kerr parameter ($\chi$) by a very unique relation, namely $q_2 = -\chi^2$.
There are on-going efforts to test this relation with the help of  GW observations \cite{krishnendu17}, X-ray binary \cite{bambi11}, black hole images \cite{johannsen10} and by timing BH-pulsar systems in the SKA era and observing stars orbiting galactic center BH during the TMT era.

\begin{figure*}
\centering
\includegraphics[width = 0.48\textwidth]{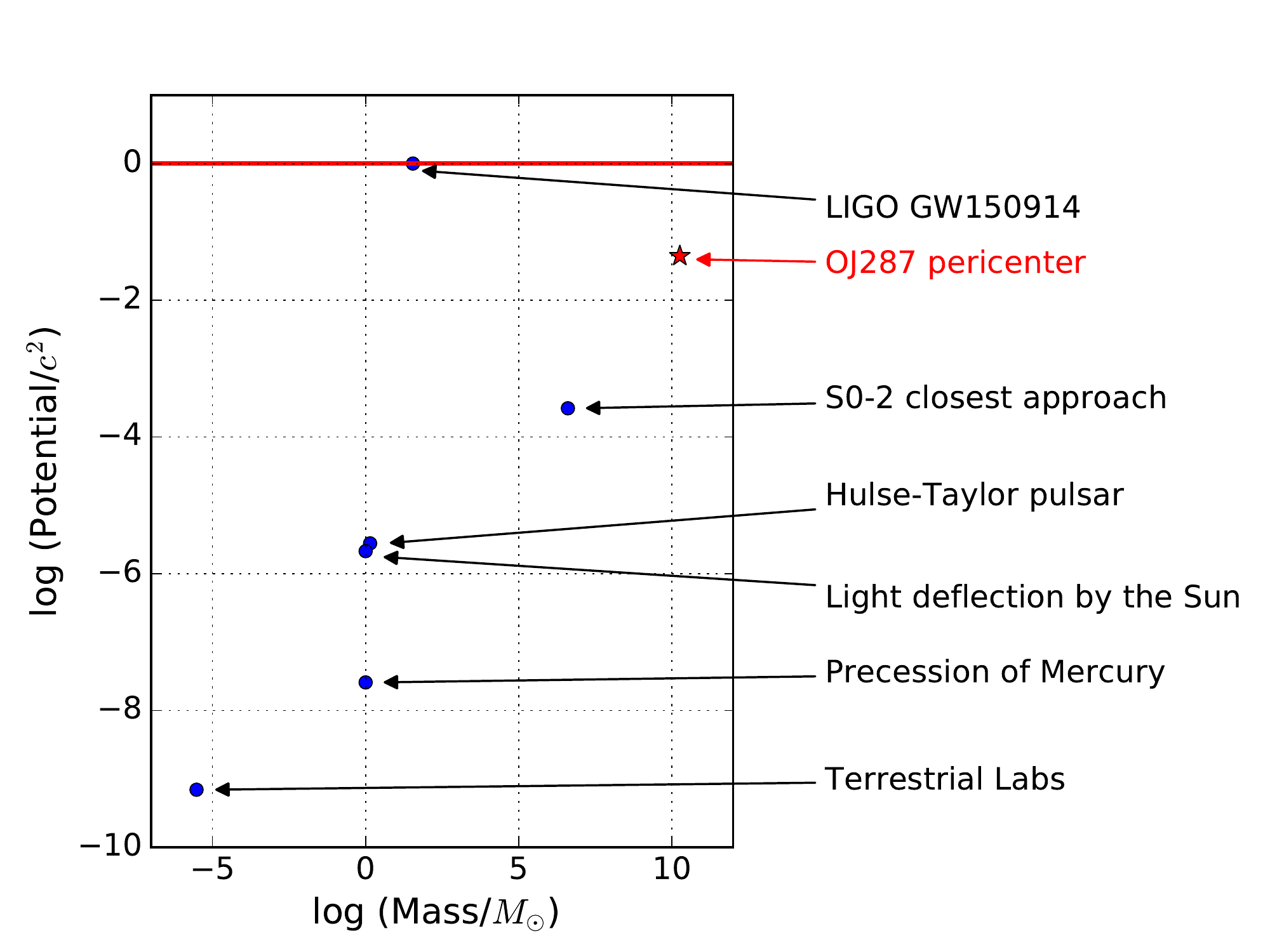}
\includegraphics[width=0.48\textwidth]{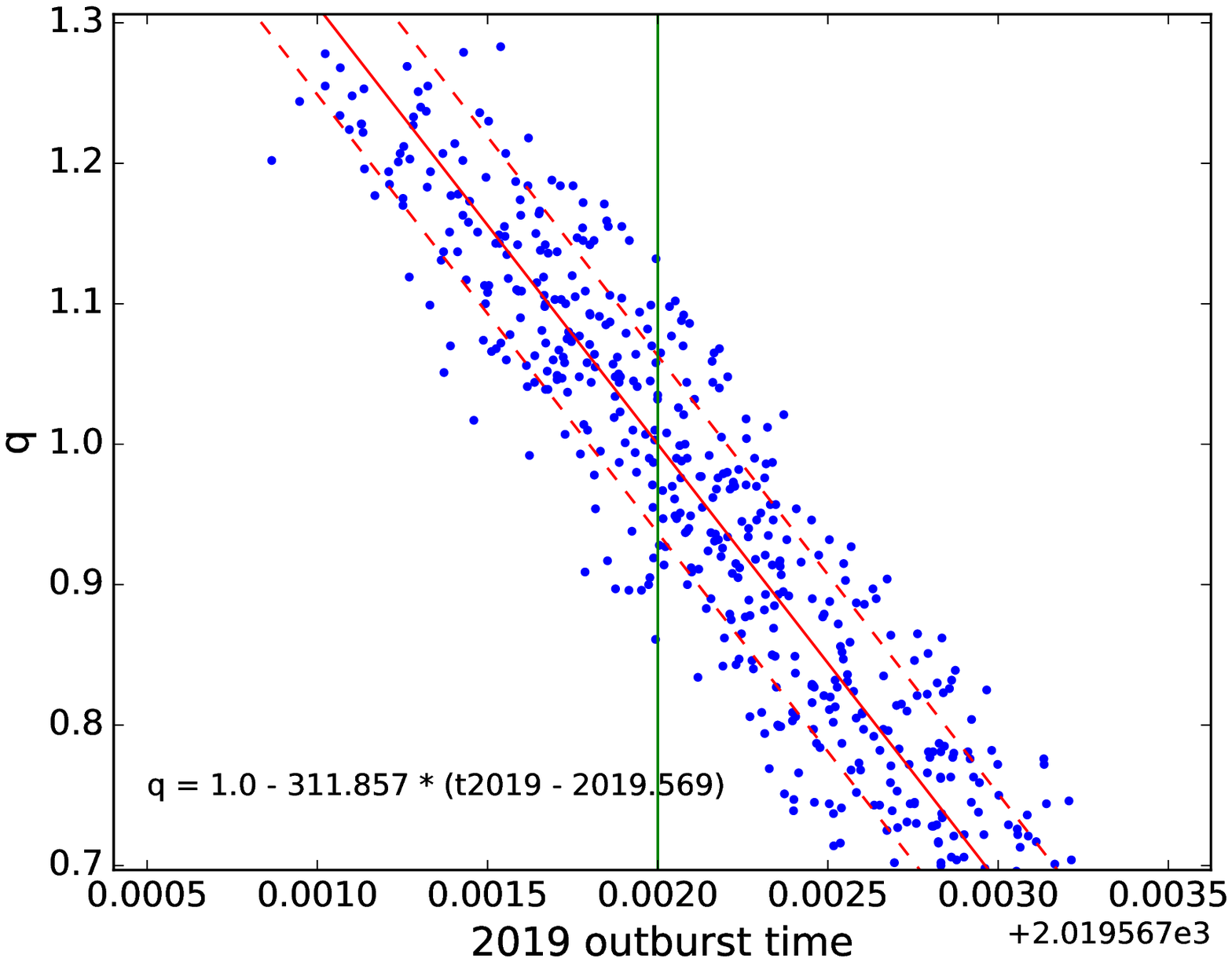}
\caption{The left panel 
plots the dimensionless gravitational potential against the gravitating mass 
in solar mass units for systems used to test Einstein's GR;
The unique position of OJ~287 is clearly visible. 
The right panel shows the correlation between the value of the $q$ parameter and the starting time of 2019 outburst where each point represents a possible solution of the binary in OJ~287. The solid red line shows our fit to the observed correlation and the dashed red lines indicate its $1\sigma$ deviation (the fit equation is given in the lower-left corner of the plot). The green vertical line represents the expected outburst starting time from the model \cite{dey18}. \textcopyright AAS 2018.}
\label{fig:no-hair_correlation}
\end{figure*}

We propose to test the above relation by replacing the primary BH quadrupole moment that appears in the $\ddot{\vek x}_Q$ term by
\begin{equation}
    q_2 = -q \, {\chi}^2\,,
    \label{eqn:no-hair}
\end{equation}
where the minus sign reflects the oblate nature of a rotating BH and in GR $q\equiv 1$ \cite{val11}.
The plan is to extract the value of the free parameter $q$ from the observed impact flare timings and at present, we can constrain the value of $q$ to be $1 \pm 0.5$.
Interestingly, an accurate determination of the starting time of the next impact flare should allow us to constrain the $q$ value at the 10\% level.
This is possible because the starting time of the expected 2019 outburst is highly correlated with the $q$ value \cite{dey18}.
In the right panel of Figure~\ref{fig:no-hair_correlation}, we show the correlation between the parameter $q$ and the starting time of 2019 outburst and the outburst will start on July 29, 2019, if the theorem holds.

Unfortunately, OJ~287 will be very close to the Sun at the predicted time and it is impossible to observe it from any ground-based observatory. 
The best strategy is to observe OJ~287 by employing space-based telescopes which are located far from the Earth during late July/early August 2019.
Indeed, we have an approved warm Spitzer Space Telescope proposal for observing OJ~287 during the epoch when the impact flare magnitude is expected to peak.
We have argued that the expected 2019 outburst light curve should be similar to that of 2007 \citep{dey18}.
This opens up the possibility of determining the epoch of the rising part of this outburst even if we miss its actual observation.
In the next section, we outline a new method 
to describe the dynamics of the BBH in OJ~287.

%%%%%%%%%%%%%%%%%%%%%%%%%%%%%%%%%%%%%%%%%%
\section{ Describing  OJ~287’s BBH dynamics via GW phasing prescription for eccentric binaries}
\label{sec:phasing}

The prospects of pursuing strong field tests of GR prompted us to develop an improved prescription to track the secondary BH trajectory in an accurate and computationally inexpensive manner.
This is a crucial requirement as we plan to employ Bayesian methods to test GR using OJ~287 observations. 
Recall that the Bayesian approach demands computation of the likelihood function millions of times that involves a similar number of BH trajectory computations. 
At present, such large scale computations are not viable as we blindly solve Equation~\ref{eqn:eom} numerically for following the secondary BH which is computationally expensive.
We adapt the GW phasing approach, detailed in Refs.~\cite{dgi04,kg06}, that provides accurate orbital phase evolution for compact binaries inspiraling along PN-accurate precessing eccentric orbits. 
% http://adsabs.harvard.edu/abs/2004PhRvD..70f4028D
% http://adsabs.harvard.edu/abs/2006PhRvD..73l4012K
The phasing approach focuses on the accurate temporal orbital phase evolution, a crucial requirement for constructing  GW templates for such binaries.
This feature is crucial for making accurate predictions for the impact flare timings as these flares occur when the secondary crosses the accretion disk of the primary at constant orbital phase angles like $0, \pi, 2\,\pi,..$.
Therefore, accurate BBH orbital phase determination should lead to precise predictions for the impact flare epochs.

 The GW phasing formalism provides an efficient way of 
 solving  both the conservative and reactive contributions to the orbital dynamics,
specified by the first and second lines in Equation~\ref{eqn:eom}.
It turns out that the 3PN-accurate conservative part of this orbital dynamics is integrable 
and admits a Keplerian-type
parametric solution as detailed in Refs.~\cite{MGS04,THG16,mikkola87}.
The existence of such a 3PN-accurate Keplerian-type parametric solution 
allows us to express the orbital phase as
\begin{equation}
\phi(t) = \lambda + W(u(l), {\cal E}, {\cal J})\,,
\end{equation}
where $u$ and $l$ are the eccentric and mean anomalies 
of the PN-accurate Keplerian parametrization, while ${\cal E}$ and 
${\cal J}$ stand for the conserved orbital energy and the angular momentum, respectively. 
The split of the angular variable $\phi$ ensures that the contributions due to  periastron 
advance remain linear in $l$ while the 
$2\pi$-periodic $W(u)$ part analytically models 
orbital time scale variations in $\phi$.
These considerations allow us to write $\lambda = ( 1+k)\,l$, where 
$k$ provides the dimensionless rate of periastron advance per orbit.
An additional equation is required to specify how $u$ varies with $l$ 
which allows us to model explicitly the  temporal evolution of $\phi$
as $l =  n ( t- t_0)$ where $n$ is the mean motion.
In practise, we solve 
 3PN-accurate Kepler equation  to find $u(l)$ and this equation 
can be symbolically written as
\begin{align}
\label{Eq_2:KE_0123}
l &=  u - e_t\, \sin u + l_{\rm 3} (u, {\cal E}, {\cal J})\,.
\end{align}
In this equation.
$l_{\rm 3}$ stands for the 3PN order corrections to the usual Newtonian Kepler equation,
namely $l = u -e_t\, \sin u$ and $e_t$ denotes 
 certain `time-eccentricity' parameter of the 
Keplerian-type parametric solution to  PN-accurate orbital dynamics.

Detailed computations reveal that the secular variations to
${\cal E}$ and ${\cal J}$ due to the addition of 
reactive contributions to the orbital dynamics, namely 
the second line in Equation~\ref{eqn:eom}, 
are identical to the PN-accurate  expressions for 
far-zone energy and angular momentum fluxes.
This is a highly desirable result as 
the far-zone energy and angular momentum fluxes are available to 
higher PN orders.
This allowed us to model orbital dynamics of compact binaries 
inspiraling under the influence of GW emission at the 2PN order 
while moving along 3PN-accurate eccentric orbits, as described 
by the first and second line contributions in Equation~\ref{eqn:eom}.
Further, the approach allowed us to tackle the spin effects  separately. 
In practise, we employ the mean motion $n$ and $e_t$ to characterize 
the orbit in place of ${\cal E}$ and ${\cal J}$ and 
GW phasing approach provides differential equations for $n$ and $e_t$.
At present, we include all the relevant 3PN (relative) contributions to the above two differential equations with the help of Refs.~\cite{klein18,arun09}.
Additionally, the approach provides  differential equations to describe
the secular evolution to $\lambda$ and $l$.
Therefore, we solve the resulting 
four ordinary coupled differential equations along with PN-accurate 
Kepler equation to obtain $u$ as a function of time.
This allows us to obtain an accurate and efficient way of describing temporal 
evolution to $\phi$, caused by various PN contributions in Equation~\ref{eqn:eom}.

It is now straightforward to compute the time when the secondary BH crosses the accretion disk of the primary ($t_{\rm imp}$).
This is obtained by finding out the epochs at which $\phi$ becomes an integer multiple of $\pi$ like $0,\ \pi,\ 2\pi,\ 3\pi,\ \dots $. 
This is justifiable as we let the accretion disk plane to be the sky plane and define the ascending node as the point where secondary BH crosses the accretion disk plane.
After gathering the impact epoch, we invoke Equation~\ref{eqn:t_outburst} to calculate the actual starting time of the associated outburst which we can compare with observational data given in Table~\ref{tab:outburst_times}.
This way we can model the observational data with the BBH central engine model for OJ~287 while adapting the GW phasing approach.

Few comments are in order. Note that when we obtain the secondary BH trajectory by numerically integrating Equation~\ref{eqn:eom}, we have to solve three complicated second order differential equations for three components of the position vector ($\vek{x}$) which is equivalent to solving six first order equations.
However, we solve only four first-order orbital averaged differential equations in the phasing approach.
Additionally, we can use larger time step for the integration of these four equations as the temporal evolution for  $n$ and $e_t$ occurs slowly on a GW radiation reaction time scales of few thousand years.
However, the position vector ($\vek{x}$) changes rapidly on an orbital time scale while numerically integrating Equation~\ref{eqn:eom}.
This forces us to use finer time steps for the integration. 
These considerations ensure that accurate and efficient computation of secondary BH trajectory when we adapt GW phasing approach to deal with BBH central engine in OJ~287.
A manuscript that details the above approach and 
 an associated publicly available code is in preparation.

%%%%%%%%%%%%%%%%%%%%%%%%%%%%%%%%%%%%%%%%%%
\section{ Summary}
\label{sec:discussion}

The bright (and unique) blazar OJ~287 is a natural candidate for hosting an SMBBH at its center. Its optical LC shows quasi-periodic doubly peaked high brightness flares with a period of $\sim 12$ years which leads to the BBH central engine model for OJ~287. This model accurately predicted the starting times of 2007 and 2015 outbursts. 
At present, we employ a sophisticated PN-accurate orbital description and a detailed impact model to calculate the binary orbit and to predict the epochs of future outbursts. The next flare is predicted to peak around July 31, 2019.
The predictive power of our BBH central engine model can be used to test different aspects of GR. 
A decade ago, our model provided the first indirect evidence for the  emission of GWs from black hole binaries  prior to the monumental direct detection of GWs from merging BH binaries by the LIGO-Virgo consortium.
Additionally, the BBH central engine model provided the test of the BH no-hair theorem and precise measurement of the starting time of the 2019 outburst will allow us to the test no-hair theorem at $\sim 10\%$ level.
We briefly presented an accurate and computationally efficient method to track the secondary BH trajectory. This is being incorporated into a Bayesian framework to fit the observational data  with the BBH central engine model. These developments should allow us to pursue unique strong field tests of GR with OJ~287 observations. 
A description of alternative models of OJ~287 is also given in the Appendix.

%%%%%%%%%%%%%%%%%%%%%%%%%%%%%%%%%%%%%%%%%%
%\section{Patents}
%This section is not mandatory, but may be added if there are patents resulting from the work reported in this manuscript.

%%%%%%%%%%%%%%%%%%%%%%%%%%%%%%%%%%%%%%%%%%
\vspace{6pt} 

%%%%%%%%%%%%%%%%%%%%%%%%%%%%%%%%%%%%%%%%%%
%% optional
%\supplementary{The following are available online at \linksupplementary{s1}, Figure S1: title, Table S1: title, Video S1: title.}

% Only for the journal Methods and Protocols:
% If you wish to submit a video article, please do so with any other supplementary material.
% \supplementary{The following are available at \linksupplementary{s1}, Figure S1: title, Table S1: title, Video S1: title. A supporting video article is available at doi: link.}

%%%%%%%%%%%%%%%%%%%%%%%%%%%%%%%%%%%%%%%%%%
\authorcontributions{L.D., A.G., M.V, P.P., and A.S. developed various facets of the astrophysical model while others contributed with observational efforts. 
All authors discussed the results and commented on various aspects of the manuscript.}

%%%%%%%%%%%%%%%%%%%%%%%%%%%%%%%%%%%%%%%%%%
%\funding{Please add: ``This research received no external funding'' or ``This research was funded by NAME OF FUNDER grant number XXX.'' and  and ``The APC was funded by XXX''. Check carefully that the details given are accurate and use the standard spelling of funding agency names at \url{https://search.crossref.org/funding}, any errors may affect your future funding.}

%%%%%%%%%%%%%%%%%%%%%%%%%%%%%%%%%%%%%%%%%%
\acknowledgments{L.D. thanks FINCA and acknowledges the hospitality of University of Turku, Finland. S.Z. acknowledges NCN grant No. 2018/29/B/ST9/0793.}

%%%%%%%%%%%%%%%%%%%%%%%%%%%%%%%%%%%%%%%%%%
\conflictsofinterest{The authors declare no conflict of interest.} 

%%%%%%%%%%%%%%%%%%%%%%%%%%%%%%%%%%%%%%%%%%
%% optional
\abbreviations{The following abbreviations are used in this manuscript:\\

\noindent 
\begin{tabular}{@{}ll}
GW & Gravitational Wave \\
SMBBH & Supermassive binary black hole \\
BBH & Binary black hole \\
SMBH & Supermassive black hole \\
AGN & Active galactic nucleus \\
LC & Light curve \\
GR & General Relativity \\

\end{tabular}}

%%%%%%%%%%%%%%%%%%%%%%%%%%%%%%%%%%%%%%%%%%
%% optional
\appendixtitles{yes} %Leave argument "no" if all appendix headings stay EMPTY (then no dot is printed after "Appendix A"). If the appendix sections contain a heading then change the argument to "yes".
\appendixsections{one} %Leave argument "multiple" if there are multiple sections. Then a counter is printed ("Appendix A"). If there is only one appendix section then change the argument to "one" and no counter is printed ("Appendix").
\appendix
\section{Alternative models for OJ~287}
\label{app:alt_models}
In what follows, 
we briefly discuss various alternative models proposed for OJ~287 over the years, how those models tried to explain the observed variability in OJ~287 and what are their shortcomings.
We begin by a scenario where a companion black hole orbits the primary inside the accretion disk of the primary, detailed in \citet{sillanpaa88}.
It produces a periodic sequence of outbursts via increased accretion flow into the primary. The typical time scale of the rise of the flare is about one month, in contrast to the observed one-day time scale. 
This is a viable mechanism for explaining the variations in the background level of OJ~287, as was shown first by \citet{sun97}. 
Another mechanism suitable for explaining background level variations is a model first proposed by \citet{katz97}. In this model, 
the secondary influences the orientation of the accretion disk, which is presumably connected to the jet in a way that causes a jet wobble in tune with the disk variations. 
This provides a good description for the background level variation in the time scale of decades \cite{val13}. However, attempts to connect this mechanism with flares have failed. 
For example, the similar kind of precessing jet model of \citet{britzen18} explains the timing of only two of the ten brightest optical flares in the historical light curve of OJ~287. The other eight bright flares have occurred during the periods when OJ~287 should have been in a quiet state according to this model. 
Therefore, its rate of success is no better than being accidental. 
We note that the model predicted 6 periods of activity lasting about 4 years each
over the last 140 years observed time span  The random rate of $24/140 \sim 1/6$ is not very different from $1/5$, the success rate of this model. 
Additionally, the model predicts roughly one year for the flux rise which is very 
different from the observed one day time scale.

\citet{valtaoja00} proposed a hybrid model 
which requires 
the secondary impacts of the accretion disk to produce 
a flare only at every other disk crossing. They assume that the disk is sharply truncated between the two impacts points, and only the closer impact produces the flare. On the other hand, they assume that the more distant crossings should be more efficient in generating tidal responses in the disk than the closer impacts. This is clearly 
counter-intuitive
to what is generally accepted about the tidal responses. In order that the sharp disk truncation would help to create two kinds of flares, the impact flare and the tidal flare, \citet{valtaoja00} have to assume that the major axis of the binary does not precess which puts an unrealistically low upper limit on the primary BH mass. If there were precession, it would gradually remove both impact points beyond the truncation radius, or bring both impacts points inside the truncation radius, depending on where the truncation happens exactly. 
Note that even a few degrees of orbital precession should have 
observational implications, the model demands periastron advance to be 
fraction of a degree per orbital revolution. This limits 
the primary mass to be $M \sim 2\times10^{8} M_{\odot}$
and the secondary black hole mass has to be very similar to 
the above estimate 
in order to produce the impact flares of the required magnitudes \citep{lv96}.
In other words, the model essentially requires an comparable mass BH binary in OJ~287.
Interestingly, 
The accretion disk in such a system is unstable within a few orbital revolutions \citep{val12} which contradicts the fact that OJ~287 has been observed over 100 years.

Interestingly, there exists a constant orbital period BH binary model where the rms prediction error has beenaround 1.5 years \citet{valtaoja00}.
A good illustration of this feature  was its prediction that a major flare should occur in the fall of the year 2006. In fact the latter part of 2006 was the time of a major dip in the light curve indicating that OJ~287 was exceptionally quiet. The major flare happened a year earlier, at the time predicted by an earlier version of our binary black hole model\citet{sun97}.
This model predicted the whole LC of OJ~287 from the year 1900 to the year 2030, and so far it has been spectacularly successful \citep{val11b}. 
In contrast, there exists a class of models which ignores the fact that the flare times are predictableeven though they have been predicted with the rms accuracy of 16 days since 1988 \citep{val18}.
The unpredictability of OJ~287 have advocated by \citet{villforth10} and more recently by \citet{qian19}. Besides being predictable, OJ~287 displays two clear periodicities of roughly 12 and 60 years when one performs a regular  Fourier analysis of its light curve \citep{val06a}. 
Note that \citet{goyal18} did not detect such periodicities and this is attributable to the use of plate calibration curves (that is, they picked the observations only when OJ~287 was just above the plate limit) instead of the OJ~287 historical light curve. Indeed, this effort showed  that the 12 and 60 yr cycles are not due to the changing plate material or choice of exposure times in the historical plate archives.
We note that quasi-periodic oscillations with this time structure were eliminated by \citet{val11b} at a high significance level.

In summary, we infer that various alternative models can explain  some of the observational aspects of OJ~287 and usually fail to explain many other features. In contrast, the BBH impact model of \citet{lv96} explains most of the features present in the optical LC of OJ~287 like, sharp rise of flares, decrease in degree of polarisation during flares, the long term periodic variations etc. This model can also explain the features present in high frequency radio observations of the jet of OJ~287 \citep{val12b,val13,dey19}. 
Most importantly, our model has accurately predicted the starting times of the widely observed 1995, 2005, 2007 and 2015 outbursts \cite{lv96, val08b, val16, val17} which places the model in much stronger ground than any of the alternative models. 
Let us note that the 1994 flare was also predicted successfully by \citet{sillanpaa88}, based on their constant period BBH model. This idea has not worked for the later flares as they are clearly not periodic. In fact,  the constant period model predicted a flare during 2018, while OJ~287 actually went to one of its deepest minimum flux values during that year. 
However, many flares have also been discovered in the historical light curve since the completion of the impact model in early 1995 \citep{val96}. They all follow the expected pattern, and in this sense there have been many more successful predictions than what was listed earlier. The most prominent of the recently discovered flares is the 1957-1959-1964 triple, of which only the middle one was known in 1995 when the impact model was first presented. The triple occurs when the major axis of the binary is lined up with the disk plane. It happens about every 60 years.
Interestingly, we are at present in the middle of the 2015-2019-2022 triple. Such triples do not exist in other models proposed so far.

%%%%%%%%%%%%%%%%%%%%%%%%%%%%%%%%%%%%%%%%%%
% Citations and References in Supplementary files are permitted provided that they also appear in the reference list here. 

%=====================================
% References, variant A: internal bibliography
%=====================================
\reftitle{References}

% The following MDPI journals use author-date citation: Arts, Econometrics, Economies, Genealogy, Humanities, IJFS, JRFM, Laws, Religions, Risks, Social Sciences. For those journals, please follow the formatting guidelines on http://www.mdpi.com/authors/references
% To cite two works by the same author: \citeauthor{ref-journal-1a} (\citeyear{ref-journal-1a}, \citeyear{ref-journal-1b}). This produces: Whittaker (1967, 1975)
% To cite two works by the same author with specific pages: \citeauthor{ref-journal-3a} (\citeyear{ref-journal-3a}, p. 328; \citeyear{ref-journal-3b}, p.475). This produces: Wong (1999, p. 328; 2000, p. 475)

%=====================================
% References, variant B: external bibliography
%=====================================
%\externalbibliography{yes}
%\bibliography{your_external_BibTeX_file}

%%%%%%%%%%%%%%%%%%%%%%%%%%%%%%%%%%%%%%%%%%
%% optional
%\sampleavailability{Samples of the compounds ...... are available from the authors.}

%% for journal Sci
%\reviewreports{\\
%Reviewer 1 comments and authors’ response\\
%Reviewer 2 comments and authors’ response\\
%Reviewer 3 comments and authors’ response
%}

%%%%%%%%%%%%%%%%%%%%%%%%%%%%%%%%%%%%%%%%%%
\end{document}